\documentclass{ws-ijmpe}
\usepackage[super,compress]{cite}

\def\lesssim{\;\raise0.3ex\hbox{$<$\kern-0.75em\raise-1.1ex\hbox{$\sim$}}\;}
\def\gtrsim{\;\raise0.3ex\hbox{$>$\kern-0.75em\raise-1.1ex\hbox{$\sim$}}\;}
\def\mdens{{\rm g~cm^{-3}}}
\def\bdens{{\rm fm^{-3}}}

\def\od{\dot\omega}
\def\pb{P_{\rm b}}

\def\msol{M_\odot}
\def\msun{M_\odot}

\begin{document}

\markboth{N. Chamel, P. Haensel, J.~L. Zdunik, A.~F. Fantina}{On the Maximum Mass of Neutron Stars}

\catchline{}{}{}{}{}

\title{On the Maximum Mass of Neutron Stars}

\author{N. Chamel}

\address{Institut d'Astronomie et d'Astrophysique,
Universit\'e Libre de Bruxelles - CP226, 1050 Brussels,  Belgium\\
nchamel@ulb.ac.be}

\author{P. Haensel}

\address{Astronomical Center, Polish Academy of Sciences,
Bartycka 18, PL-00-716 Warszawa, Poland\\
haensel@camk.edu.pl}

\author{J.~L. Zdunik}

\address{Astronomical Center, Polish Academy of Sciences,
Bartycka 18, PL-00-716 Warszawa, Poland\\
jlz@camk.edu.pl}

\author{A.~F. Fantina}

\address{Institut d'Astronomie et d'Astrophysique,
Universit\'e Libre de Bruxelles - CP226, 1050 Brussels,  Belgium\\
afantina@ulb.ac.be}

\maketitle

\begin{history}
\received{Day Month Year}
\revised{Day Month Year}
\end{history}

\begin{abstract}
One of the most intriguing questions about neutron stars concerns their maximum mass.
The answer is intimately related to the properties of matter at densities far beyond that
found in heavy atomic nuclei. The current view on the internal constitution of neutron
stars and on their maximum mass, both from theoretical and observational studies, are briefly
reviewed.
\end{abstract}


\ccode{PACS numbers:04.40.Dg; 26.60.Kp; 97.10.Nf; 97.60.Jd}


\section{Introduction}
\label{sect:introd}

Neutron stars (NSs) are the densest stars observed in the Universe, with average density
exceeding significantly the normal nuclear density $\rho_0=2.8\times 10^{14}~\mdens$ found
in heavy nuclei and corresponding to the baryon number density $n_0=0.16~\bdens$.
They are observed as various astrophysical sources like radio and X-ray pulsars, X-ray bursters,
compact thermal X-ray sources in supernova remnants, rotating radio transients; they are also promising
sources of gravitational waves.

The structure of a NS is determined by the equation of state (EoS) of dense matter, i.e.
the relation between the matter pressure $P$ and the mass density $\rho={\mathcal E}/c^2$ where
$\mathcal{E}$ is the energy density and $c$ the speed of light (for a detailed review of
the EoS and the structure of a NS, see e.g. Ref.\refcite{haen07}). A remarkable consequence of
the general theory of relativity is the existence of a maximum NS mass $M_{\rm max}$.
The evolution of ideas related to the origin of this limiting mass are briefly discussed in
Sec.~\ref{sect:Mmax.origin}.

The actual value of $M_{\rm max}$ depends on the EoS and therefore on the internal structure
of NSs. In spite of their names, NSs are not only made of neutrons. With densities ranging from a
few $\mdens$ as in ordinary matter up to about $10\rho_0$, the interior of a NS is characterized
by very different phases of matter, either homogeneous or inhomogeneous. Our current view on the
constitution of a NS as well as the corresponding EoS are reviewed in Sec.~\ref{sect:EOS.modern}.

The knowledge of the maximum mass of compact stars has important consequences for identifying 
compact astrophysical sources: those with a mass lying below the limiting mass are compact 
stars, while the others have to be black holes. Due to the uncertainties in the values of 
$M_{\rm max}$, the nature of some objects, especially soft X-ray transients, remains elusive. 
This is particularly the case for GRO~J0422$+$32, whose measured mass $3.97 \pm 0.95~\msun$~\cite{gel03}
($\msun$ being the mass of the Sun)
suggests that it is a stellar black hole. However, it has been recently argued that the mass of this object 
(as well as that of other similar sources previously identified as black holes) could be substantially lower 
due to systematic errors~\cite{kre12}.

Even though the EoS of NS cores still remains very uncertain, an upper bound on the NS 
mass can be inferred from general considerations, as reviewed in Sec.~\ref{sect:Mmax}.
The impact of rotation on the maximum NS mass is discussed in Sec.~\ref{sect:Mmax.rot}.

While a reliable  theoretical calculation of the maximum mass is extremely difficult, measurement
of NS masses can provide solid observational (in terrestrial laboratory physics one would say ``experimental'')
constraints on dense matter theories.
The most precise measurements of NS masses in binary systems are reviewed in Sec.~\ref{sect:observations}.

\section{The origin of the maximum mass}
\label{sect:Mmax.origin}
\subsection{A prelude: the maximum mass of white dwarfs}
\label{sect:chandra.mass}
The existence of a limiting mass for degenerate stars was first discovered
in the case of white dwarfs (WDs). The exact calculation of the maximum mass
$M^{\rm WD}_\mathrm{max}$ was carried out by Chandrasekhar~\cite{Chandra1931}
within Newtonian gravitation theory.
 \footnote{Earlier estimates of $M_{\rm max}^{\rm WD}$ were given by
Anderson~\cite{Anderson1929} and Stoner~\cite{Stoner1930}. The history
of $M_{\rm max}^{\rm WD}$ is described, e.g., in Ref.~\refcite{Shaviv2010}.}
Chandrasekhar considered non-rotating WDs built of a completely ionized plasma
of nuclei with $Y_e$ electrons per nucleon. He treated electrons as
an ideal Fermi gas and assumed that nuclei do not contribute to pressure.
The now so called {\it Chandrasekhar mass} limit, $M_{\rm Ch}$, results from
the fact that electrons become relativistic for $\rho \gg \rho_e$ with
$\rho_e=m/\lambda_e^3 \sim 10^7~\mdens$ where $\lambda_e$ is the electron
Compton wavelength and $m$ denotes the average mass per electron. However, 
special relativity limits the maximum stiffness
of the electron gas due to the effect that the increase in pressure with increasing
density cannot exceed $dP/d\rho=c^2/3$ where $c$ is the speed of light.
For central density $\rho_{\rm c}\longrightarrow +\infty$, the WD mass thus tends
asymptotically to the upper limit
\begin{equation}
M_{\rm Ch}=1.46\;\left(2Y_e\right)^2~\msun~,
\label{eq:M.Ch}
\end{equation}
where $Y_e$ denotes the lepton fraction.
Later, and independently of Chandrasekhar, Landau calculated the value of the
maximum mass of a degenerate star~\cite{Landau1932}.
\footnote{Actually, Landau did not mention WDs in his paper
and considered the general case of stars built of dense degenerate matter.}
He showed that hydrostatic equilibria of stars supported by the pressure of degenerate
electrons only exist for $M<M_{\rm L}$ with
\begin{equation}
M_{\rm L}={3.1\over m^2}\left(\hbar c\over G\right)^{3/2}=1.5\;\left(2Y_e\right)^2~\msun~,
\label{eq:M.Land}
\end{equation}
in which $\hbar$ the Dirac's constant and $G$ the gravitational constant. Landau suggested that
stars having a mass $M>M_{\rm L}$
would collapse thus ``forming one gigantic nucleus'' (by ``nucleus'' he meant an atomic
nucleus). This description has often been considered as an anticipation or even a prediction of NSs.
\footnote{Landau's paper was actually written and submitted for publication {\it before}
the discovery of the neutron! See Ref.~\refcite{Yakovlev2012} for a review of the history of
NS in the 1930s and the role of Lev Landau.}
\subsection{The maximum neutron-star mass from Landau's method}
\label{sect:landau.mass}
Landau~\cite{Landau1932} derived the maximum mass of a WD arguing that the hydrostatic
equilibrium of a degenerate star corresponds to a minimum of its total energy. It is
straightforward to adapt this reasoning to a Newtonian model of NSs. Let us consider a
self-gravitating sphere of radius $R$ and total mass $M$ containing $N$ degenerate
neutrons with mass $m_n$. At sufficiently high densities, neutrons become relativistic. Neglecting the
interaction energy between neutrons, the internal energy of the star is estimated from
the Fermi energy $\varepsilon_{\mathrm{F}n}$ of ultra-relativistic neutrons,
\begin{equation}
    \varepsilon_{\mathrm{F}n}\simeq \hbar c
    \left({ N /R^3}\right)^{1/3}~,\qquad
     E_\mathrm{int}(N,R)\simeq N \varepsilon_{\mathrm{F}n}\simeq
    (\hbar c / R)\, N^{4 / 3}~.
\end{equation}
On the other hand, the gravitational energy of the star is given by
\begin{equation}
    E_\mathrm{grav}(N,R)\simeq -{G M^2 / R} =
    -{G N^2 m_n^2 / R}~.
\end{equation}
The total energy thus takes the form $E=E_\mathrm{int}+E_\mathrm{grav}=\alpha/R$,
where $\alpha$ depends on $N$ but is independent of $R$. If $\alpha<0$, the
equilibrium configuration corresponds to $R\rightarrow 0$. Therefore stable stars
can only exist if $\alpha >0$, or equivalently
\begin{equation}
M<\left(\frac{\hbar c}{G m_n^2}\right)^{3/2} m_n \approx 1.8~\msun~.
\label{struct-Landau.Mmax}
\end{equation}
This derivation is based on two assumptions:
(i) dense matter consists of an ideal Fermi gas of ultrarelativistic neutrons
(neutrons are therefore supposed to exist at densities $\rho\gg 10^{15}~\mathrm{g~cm^{-3}}$!),
(ii) NSs can be treated by Newton's theory of gravitation.
Both assumptions are unrealistic. Therefore, a reasonable value of the maximum mass
is just a lucky coincidence. The crucial effect of general relativity will be reviewed in the
next section.
\subsection{General relativity and the existence of a maximum mass}
\label{sect:GR.mass}

With a mass comparable to that of the Sun and a radius of about $10$~km, NSs are
extremely compact objects: the Schwarzschild radius, defined by $r_\mathrm{g}=2 G M/c^2$,
represents a sizable fraction of the star's radius
$R$ whereas for all other stars $r_\mathrm{g}\ll R$ (the limit $r_\mathrm{g}=R$ is only reached for black
holes). A realistic description of NSs must therefore rely on Einstein's theory of general
relativity.

It is generally assumed that the interior of a NS is made of cold catalyzed matter at the
end point of thermonuclear evolution, i.e. matter in full thermodynamic equilibrium at zero
temperature and zero entropy~\cite{htww65}. This assumption implies that the stress-energy
density tensor of NS matter is that of a perfect fluid, as shown in Chap. 9 of Ref.~\refcite{htww65}. 
Indeed, if shear stresses existed in the star, the star would not be in full equilibrium. The cold-catalyzed matter hypothesis
thus greatly simplifies the determination of the NS structure. Of course, a real NS may
sustain shear stresses in its solid crust and possibly in its core (see, e.g., Sec. 7.7 in Ref.\refcite{haen07}). However, these
stresses are presumably very small since the interior of newly-born NSs is expected to be a
very hot liquid (on the other hand, stresses might be induced by the presence of a magnetic field).

Following the same line of reasoning, it can be shown~\cite{htww65} that the mass density $\rho=\mathcal{E}/c^2$,
$\mathcal{E}$ being the total energy density, can only depend on the baryon density $n$. The
pressure $P$ is then also completely determined by $n$ and is given by~\cite{htww65}
\begin{equation}
P=n^2 \frac{d(\mathcal{E}/n)}{dn}\, .
\end{equation}

We will further assume that the star is static and spherically symmetric. The Tolman-Oppenheimer-Volkoff
(TOV) equations~\cite{tol39,ov39} of hydrostatic equilibrium are given by
\begin{equation}
\frac{dP}{dr}=-\frac{G\rho \mathcal M}{r^2}\left(1+\frac{P}{\rho
c^2}\right)\left(1+ \frac{4\pi P r^3}{\mathcal M c^2}\right)\left(1-\frac{2G
\mathcal M}{r c^2}\right)^{-1}\, ,\label{tov1}
\end{equation}
where the function $\mathcal M(r)$ is defined by
\begin{equation}
\frac{d\mathcal M}{dr}=4\pi r^2 \rho\, ,\label{tov2}
\end{equation}
with the boundary condition $\mathcal M(0)=0$. In order to solve these equations,
an EoS, i.e. a relation between the pressure $P$ and the mass density $\rho$, must be specified.
The function $P(\rho)$ depends on the properties of dense matter which still remain very uncertain
in the core of NSs. However, a few general assumptions can be made.
\begin{itemize}
\item In the absence of any evidence to the contrary, gravity is always attractive so that
the mass density must be positive :
\begin{equation}\label{attrac-grav}
\rho\geq 0\, .
\end{equation}
\item In order for the NS matter to remain locally in an equilibrium state, it must
be stable against contraction (Le Chatelier's principle). Therefore the function $P(\rho)$
must satisfy the following constraint:
\begin{equation}\label{micro-stability}
\frac{dP}{d\rho}\geq 0 \, .
\end{equation}
Since the pressure of ordinary matter is positive, this condition also implies that the pressure
remains positive at the higher densities prevailing in NSs:
\begin{equation}\label{pos-pressure}
P\geq 0 \, .
\end{equation}
\item The condition that the sound speed does not exceed the speed of light reads~\cite{curt50}
\begin{equation}\label{causality}
\frac{dP}{d\rho}\leq c^2 \, .
\end{equation}
\end{itemize}
This inequality is generally considered as a condition stemming from Lorentz invariance and causality.
However, as explained, e.g., in Ref.~\refcite{ell07} and in Chap. 11 of Ref.~\refcite{Fayngold2008}, the actual
situation is not so simple.

Conditions~(\ref{attrac-grav}) and (\ref{pos-pressure}) imply that $2G \mathcal M(r)/(r c^2)<1$ everywhere
inside the star~\cite{bondi64}. As a consequence, the pressure inside the star is decreasing
outwards and vanishes at the surface. The structure of the star can thus be obtained by
integrating Eqs.~(\ref{tov1}) and (\ref{tov2}) from the center with a given central pressure
$P(r=0)=P_c$ out to the radial coordinate $R$ (the circumferential radius of the star) for which
$P(r=R)=0$. The gravitational mass of the star is then given by $M\equiv \mathcal M(R)$. 
It is Zwicky~\cite{zwi38,zwi39} who first pointed out that this gravitational mass should be
distinguished from the baryon or rest mass defined by the sum of baryon masses in the star\footnote{Zwicky referred to
the gravitational (baryon) mass as the ``effective'' (``proper'') mass.}.
The difference between these two masses is of direct astrophysical interest as it represents the energy
released during the core-collapse of massive stars in type II supernovae.

Equation~(\ref{tov1}) describes the balance between the radial gravitational pull
acting on a matter element of unit volume and the net radial pressure force acting
on it. The first factor on the right-hand side is the Newtonian expression of the
gravitational pull~\footnote{However the factor $-G \mathcal M(r)\rho/r^2$ does
not coincide with the Newtonian gravitational pull because the function $\mathcal
M(r)$ is not just the sum of the rest mass of all particles within $r$
but is defined in terms of the mass density $\rho=\mathcal{E}/c^2$, where $\mathcal{E}$ is the
macroscopically averaged energy density of matter.}. It is multiplied by three
general relativistic factors, each one amplifying the gravitational pull. The two
factors $1 + 4\pi P r^3/(\mathcal{M}c^2)$ and $1 + P /(\rho c^2)$ increase with
increasing pressure, which itself increases toward the center of the star as shown
above. The  factor $1/\left(1 - 2G\mathcal M /(r c^2)\right)$ is of a different
character and arises from the space curvature in the radial direction, generated by
the mass distribution. To support an increase of the mass $M$, an increase of the
central pressure $P_{\rm c}$ is needed. This may be achieved only by the
compression of matter, which in turn amplifies the gravitational pull due to the
increasing space curvature. This makes the increase of $M$ by the increase of
$P_{\rm c}$ harder and harder. As early as 1916, Karl Schwarzschild~\cite{schw16} published the
exact solution of Einstein's equations for a spherical star made of incompressible
matter with density $\bar\rho$ and noticed that if $P_{\rm c}\rightarrow +\infty$
then $R\rightarrow (9/8)r_\mathrm{g}$. As a consequence, there exists a maximum
mass $M_{\rm max}^\mathrm{inc}$ above which the star cannot be in hydrostatic
equilibrium. This limiting value for the mass is a direct consequence of general
relativity: there is no such limit on the mass of incompressible-fluid stars in
Newtonian gravitation. If a maximum mass exists for an incompressible fluid, then
it should exist for \emph{any} EoS of matter with finite compressibility. However,
the {\it value} of the maximum mass depends on the EoS. Fritz Zwicky, who first
speculated about the existence of NSs with Baade in 1933~\cite{bz34}, applied the
Schwarzschild's solution to estimate the maximum NS mass~\cite{zwi38,zwi39}.
Assuming that the average density in NSs is comparable to that inside heavy atomic
nuclei, i.e. $\bar\rho\simeq 10^{14}~\mdens$, he thus found for the maximum mass
$M_\mathrm{max}\simeq 11 \msun$. In 1933, Sterne~\cite{ster33} showed that for
sufficiently high densities, matter becomes more and more neutron rich due to
electron capture. In 1939, Oppenheimer and Volkoff~\cite{ov39} solved
Eqs.~(\ref{tov1})-(\ref{tov2}) considering a star containing an ideal Fermi gas of
neutrons and found a very low value for the maximum mass: $M_\mathrm{max}\simeq 0.7
\msun$. This is less than one half of the Chandrasekhar mass limit for WDs.
Their calculations thus suggested that NSs could not be formed from the collapse of
ordinary stars during supernova explosions, as proposed by Baade and Zwicky a few
years earlier~\cite{bz34}. However, as clearly pointed out by
Zwicky~\cite{zwi38,zwi39}, the interior of a NS is unlikely to contain only
neutrons. In 1946-1947, van Albada~\cite{alb47} carried out the first detailed
study of dense matter and predicted the appearance of a neutron gas at densities
$\rho\simeq 5\times 10^{11}~\mdens$. In the 1950s, Wheeler and his
collaborators~\cite{htww65} calculated the EoS of matter over the full range of
densities encountered in NSs, assuming that their core consists of free neutrons,
protons and electrons in beta equilibrium. The maximum mass they obtained was
slightly lower than that found by Oppenheimer and Volkoff due to the presence of
protons. It was later realized that nuclear forces are very strong and cannot be
ignored. Cameron first showed in 1959~\cite{cam59} that the inclusion of nuclear
forces considerably stiffens the EoS thus increasing the maximum mass to
$M_\mathrm{max}\simeq 2 \msun$. He also pointed out that the core of a NS is likely
to contain hyperons. A few years later, Ivanenko and Kurdgelaidze~\cite{ivk69}
suggested that NS cores may be made of quarks, and soon afterwards such quark stars 
were studied by Itoh~\cite{itoh70}. Despite the progress in nuclear and
particle physics, the determination of the maximum NS mass continued to be a major
issue. For instance, in 1971 Leung and Wang~\cite{lw71} argued that the mass of a
NS is unlikely to exceed $0.5~\msun$! Even though the EoS of dense matter is now
fairly well-known at densities $\rho \lesssim \rho_0$, its high-density part still
remains very uncertain.

\section{Modern equations of state  of neutron-star matter and maximum mass}
\label{sect:EOS.modern}

The interior of a NS is expected to exhibit very different phases of matter~\cite{haen07}, as
emphasized by Zwicky himself~\cite{zwi38,zwi39}. In what follows, we will briefly review the
internal constitution of a NS according to the cold catalyzed matter hypothesis, i.e. matter in
its absolute ground state~\cite{htww65}. Matter in a real NS is presumably not fully catalyzed,
especially in binary systems where a NS can accrete material from its companion. However the
deviations, which could be very large in the outermost layers of the star (see, e.g.,
Ref.~\refcite{lrr} and references therein), are not expected to significantly impact the maximum mass.
A NS has an onion-like structure (see, e.g., Figure 1.2 of Ref.~\refcite{haen07}).
Moving radially inward from the surface to the center, one encounters: the atmosphere,
the ocean, the outer crust, the inner crust, the outer core, and the inner core.
The atmosphere is a thin (typically a few cm for a thousand years old NS) gaseous
plasma layer where the spectrum of photons emitted by NSs is formed. Then comes the
few meters deep ocean of a liquid plasma (it contains less than $10^{-8}$  of the
mass of the star), followed by a solid outer crust of a crystal lattice of nuclei
immersed in an electron gas. The outer crust is a few hundred meters thick,
contains some $10^{-5}$ of the mass of a NS, the density at its bottom edge is $\simeq
4\times 10^{11}~\mdens$. The composition of the outer crust is completely determined by
experimentally measured atomic masses up to a density of about $5\times 10^{10}$~g~cm$^{-3}$
(i.e., around $200$~m below the surface for a $1.4 M_\odot$ NS with a radius of $10$~km~\cite{pgc11,wolf13}).
Beneath the outer crust lies a significantly thicker ($\sim 1-2$ km) inner crust composed of a crystal
lattice of neutron-proton clusters immersed in an electron gas and a neutron liquid (see, e.g., Ref.~\refcite{lrr} for a review).
Typically, it contains, together with the outer layers above it, about $0.01$ of the NS mass (see, e.g., Ref.~\refcite{pcgd12}).
The density at its bottom is about $\sim \rho_0/2$. The liquid core is divided into two regions: (i) an outer core with
a density ranging from $\sim \rho_0/2$ up to $\sim 2\rho_0$, and composed mostly of neutrons, with a
few percent admixture of protons, electrons and muons, and (ii) an inner core whose density could
reach $\sim 10\rho_0$. The structure and the composition of the inner core is poorly known: does it contain
nucleons only? nucleons and hyperons? quark matter? meson condensates? As far as the value of $M_{\rm max}$
is concerned, the contribution of the crust to it is so small, that the uncertainties in its EoS will not
be discussed further. On the contrary, the uncertainties related to the composition and the EoS of the inner
core play a dominant role for the value of $M_{\rm max}$, and this is what we will review in the rest of
this section.

\subsection{Nucleonic core}
\label{sect:EOS.N}
The crust dissolves into a uniform liquid when the density reaches about $\sim
10^{14}$~g~cm$^{-3}$ (about half the density found at the center of heavy nuclei).
This has motivated many studies of homogeneous and beta-equilibrated matter in
which the only hadrons are nucleons, and which is neutralized by a homogeneous
lepton gas (electrons and, at higher densities, muons). These studies consist of
simple extensions of the large number of many-body calculations performed since the
beginning of the 1950s on so-called nuclear matter, consisting of just neutrons and
protons (the Coulomb force being switched off) which interact via ``realistic''
nuclear forces fitted directly to experimental nucleon-nucleon phase shifts and to
the properties of bound two- and three-nucleon systems . The EoS of purely
nucleonic NS matter has been determined in such many-body calculations up to the
highest densities found in the most massive NSs. However, even though calculations
based on different many-body methods (see, e.g., Chap. 5 in Ref.~\refcite{haen07} for a
review) yield comparable results at densities $n\lesssim 2-3
n_0$~\cite{sam10,bb12,gan12,tews13}, there remains considerable disagreement at
higher densities~\cite{ls08,fuchs08,bm12,bprsv12,sam12}. This leads to a spread in
the predictions of the maximum mass between $1.8M_\odot$ to
$2.5M_\odot$~\cite{apr98,ls08,fuchs08,sam10}. The high-density part of the
EoS of symmetric nuclear matter, at densities between $\sim 2\rho_0$ and $\sim
4\rho_0$, can be constrained by studying the flow of matter in heavy-ion collision
experiments~\cite{dan02}. However this analysis still allows for a rather large
range of degrees of stiffness. Alternatively, measurements of the kaon and pion
productions in heavy-ion collisions~\cite{fuchs01,stu01,har06,xiao09} seem to
suggest a very soft EoS. Various exotic mechanisms such as a ``fifth
force"~\cite{wen09} or variations of the gravitational constant~\cite{wen12} have
been proposed to account simultaneously for both this result and the existence of
massive NSs such as PSR J1614$-$2230~\cite{Demorest2010} (see
Sec.~\ref{sect:observations}). On the other hand, these experiments only probe the
EoS up to a few times normal density and therefore, they do not exclude the
possibility of a strong stiffening of the EoS at the much higher densities
prevailing in NS cores. In addition, the constraints inferred from heavy-ion
collisions are indirect, in that they depend on the specific transport models used
in the analysis. Moreover, even if the uncertainties in the models can be reduced,
it is not clear that heavy-ion collisions could shed light on the properties of NS
cores since the conditions are radically different (hot matter off equilibrium in a
finite system vs cold matter in equilibrium in an essentially infinite system).

\subsection{Hyperonic inner core}
\label{sect:EOS.NH}

The inner core of a massive NS is likely to contain hyperons~\cite{haen07,glen}.
The appearance of hyperons softens considerably the EoS, as compared to the purely
nucleonic EoS. Equilibrium with respect to weak interactions implies, that the most
rapidly moving energetic nucleons are replaced by more massive, slowly moving
hyperons. Actually, the softening is so strong that it leads to a ``hyperon
puzzle''.  According to Brueckner-Hartree-Fock (BHF) calculations using realistic
two- and three-body forces~\cite{vid11,bur11,sch11}, the appearance of hyperons in
dense matter lowers the maximum NS mass to an almost unique value around $1.3-1.4
M_\odot$. To make things worse, it has been recently found that 3-body forces
cannot provide enough pressure to increase the maximum mass beyond this
value~\cite{vlppb11}. Simultaneously, some relativistic mean-field (RMF)
calculations including hyperons can support NSs as massive as PSR
J1614$-$2230~\cite{bed11,sul12,jia12,weis12,zhao12,bon12,colu13}. This discrepancy can be
understood at least partly from the fact that the maximum mass depends very
sensitively on the various hyperonic couplings, and these are determined very
poorly since the limited nuclear and hypernuclear data constrain
the EoS only in the vicinity of the saturation density, whereas the maximum NS mass
is mostly determined by the EoS at much higher densities (typically between $\sim
5\rho_0$ and $10\rho_0$). Indeed, it has been shown that to obtain $M_{\rm
max}>2\;\msun$ one has to introduce an additional high-density repulsion between
hyperons, due to the exchange of the hidden-strangeness $\phi$ meson. Moreover, it
has been shown that a specific breaking of the SU(6) symmetry relating the
vector-meson  - hyperon coupling constants to the vector-meson - nucleon ones can
rise the maximum mass of NSs with hyperonic cores well above
$2\;\msun$~\cite{weis12,colu13}. Summarizing, getting $M_{\rm max}>2\;\msun$ for NSs with
hyperon cores in a RMF model consistent with nuclear and hypernuclear experimental
data, requires a tuning of the model parameters in the hyperon sector.

\subsection{Mesonic  inner core}
\label{sect:EOS.M}

The coupling of mesons to baryons generates strong interactions in dense matter.
The mesons mediating this strong interactions are virtual. However, the
meson-baryon coupling in some two-particle states in dense matter  can be
sufficiently attractive so as to produce real mesons. As the mesons are bosons,
their ground state would correspond to a boson condensate (for a review of pion and
kaon condensation in dense baryon matter, see, e.g., Sec.7.3. and 7.4 of
Ref.~\refcite{haen07}). Hypothetical pion condensation or  kaon condensation would
soften the EoS of dense matter compared to the un-condensed state, and therefore
would be unfavorable to large value of $M_{\rm max}>2\;\msun$. Still, some RMF
models including kaon condensates are able to predict the existence of massive NSs
with $M>2\;\msun$~\cite{gup12}. Consistency of the kaon condensation model with
$2\;\msun$ pulsar necessitates, however, a tuning of the parameters of the RMF
model.

\subsection{Quark inner core}
\label{sect:EOS.BQ}

The modern fundamental theory of the structure and interactions of hadrons is Quantum
Chromodynamics (QCD). Terrestrial nuclear and hypernuclear physics involves nucleons,
hyperons, hypernuclei, and mesons. From the point of view of the QCD, it involves three
lightest quarks confined in baryons and mesons.
Due to the confinement, quarks do not need to be considered explicitly and ``effective
theory'' with baryons interacting via the exchange of mesons is sufficient. Weak interactions
(described by the Standard Model) will lead to the appearance of leptons in dense matter.
The fundamental question to be answered is this: up to what density can this effective model
be used to describe cold dense matter?

Let us consider the ``fundamental (QCD) picture'', with dense cold matter composed of
quarks and leptons. Two remarkable properties can be stated: (i) confinement of quarks
to hadrons at sufficiently low density and (ii) asymptotic freedom at sufficiently high
density (matter then behaves as a quasi-ideal Fermi gas of quarks with a very simple 
EoS\cite{CollinsPerry1975}: $P\simeq \rho c^2/3$). For intermediate densities, the matter is a plasma
of quarks interacting via the exchange of gluons. Both the value of the deconfinement density
$\rho_{\rm dec}$ and the EoS for $\rho\gtrsim\rho_{\rm dec}$ are difficult to calculate,
because interactions are very strong.
For this reason, different effective phenomenological models have been
developed leading to a large spread of predictions for the maximum NS mass (see,
e.g., Ref.~\refcite{alf07} and references therein). The uncertainties pertaining to
these calculations lie on the fact that these models lack a direct relationship
with QCD. On the other hand, perturbative QCD calculations~\cite{kur10} can predict
$M_{\rm max}\gtrsim 2~\msun$ provided the quark interactions are sufficiently
strong. But the region of the validity of the perturbative  calculations resulting
from asymptotic freedom is reached for $\rho>10^{18}~{\rm g~cm^{-3}}$ - far larger
than the maximum density expected to be found at the center of the most massive NSs,
$\rho\lesssim 5\times 10^{15}~{\rm g~cm^{-3}}$. The existence of massive NSs with quark cores 
(so called hybrid stars) and with a mass $M>2\;\msun$ requires (i) a very high stiffness of the quark 
matter EoS (i.e., the speed of sound has to be sufficiently close to $c$), (ii) a sufficiently low value of $\rho_{\rm dec}$,
and (iii) a small density jump at the hadron-quark phase transition (see, e.g., Refs.~\refcite{bon12,colu13,zh13,cfpg13}).
In turn, this can only be achieved by a very fine tuning of the quark matter model parameters.

\subsection{Strange matter and other exotica}
\label{sect:exotica}
In 1971, Bodmer~\cite{bod71} speculated that atomic nuclei do not represent the
true ground state of ordinary matter. As a consequence, atomic nuclei would
``collapse'' into very compact objects  of supranuclear density, after a time that
is sufficiently long to explain the apparent stability of ``normal'' nuclei. In
1984, Witten~\cite{wit84} showed that the true ground state of matter could consist
of quasi-free $u$, $d$ and $s$ quarks. If this hypothesis is true, some NSs could
actually be ``strange'' stars, built of a self-bound quark matter. Detailed models
of such stars were developed soon afterward (see e.g. Chap. 8 of Ref.~\refcite{haen07} for a
review). The internal structure of a strange star is expected to be very different
from that of a NS. In particular, the density at the surface of a strange star is
predicted to be huge, of order $10^{15}~\mdens$ (to be compared to a few $\mdens$
at the surface of a NS) and does not differ much from the density at the center of
the star. On the other hand, the maximum mass of these hypothetical strange stars
is found to be not much different from that of NSs. Substantially higher masses are
predicted for solid quark stars (see, e.g., Ref.~\refcite{lai09}) and for even more exotic
compact stars (Q-stars, see, e.g., Sec. 8.20 of Ref.~\refcite{haen07}). However, these models
assume an exotic state of matter at the density below $2\rho_0$ and even below
$\rho_0$ (some models of Q-stars). A very reasonable condition  that the density of
exotic matter exceeds $2\rho_0$ (i.e., that below  $2\rho_0$   dense matter is in a
normal, nucleonic state) pushes down $M_{\rm max}$ of solid-quark stars and Q-stars
to quite ordinary values $2-2.5~\msun$ (see Sec. 8.20 of Ref.~\refcite{haen07} for a more
detailed discussion of this point).

\subsection{Summary of the maximum neutron-star mass predictions}

Table~1 summarizes the predictions of various microscopic and effective 
calculations of the maximum NS mass with suitable references.

\begin{table}[pt]
\tbl{Maximum neutron-star mass as predicted by different theories of
dense matter. The core is assumed to contain nucleons (N), nucleons and hyperons
(NH),  nucleons and quarks (NQ). Microscopic calculations: Brueckner Hartree-Fock
(BHF)~\cite{ls08,vid11,bur11,sch11}, Dirac Brueckner Hartree-Fock (DBHF)~\cite{fuchs08,sam10}, 
variational chain summation method (VCS)~\cite{apr98}, perturbative quantum chromodynamics 
(pQCD)~\cite{kur10}. 
Effective models: Relativistic Mean Field (RMF)~\cite{Shen2011,weis12,colu13}, 
Nambu-Jona-Lasinio (NJL)~\cite{zh13,Blaschke2010,bon12}, Modified Bag Model
(MBM)~\cite{Ozel2010,Weissenborn2011}. If the largest maximum mass $M_{\rm max\, 2}$ for a given class of models exceeds 
$2.0 M_\odot$, and the smallest maximum mass $M_{\rm max\, 1}$ is lower than 
$2.0 M_\odot$ we present the narrower range of masses $2M_\odot - M_{\rm max\, 2}$ consistent 
with observations. If, however,  $M_{\rm max \, 2}<2.0 M_\odot$, then the range of $M_{\rm max}$ 
shown is $M_{\rm max\, 1} - M_{\rm max\, 2}$; such a class of models is ruled out by observations.
For further explanations see the text. }
 {\begin{tabular}{@{}cccccccccc @{}} \toprule
   & BHF & BHF & DBHF & VCS  & pQCD &  RMF & RMF & RMF/NJL  & RMF/MBM \\
   & (N) & (NH) & (N)  & (N) & (NQ)  & (N) & (NH) &  (NQ)   &  (NQ) \\
\colrule
$M_{\rm max}/M_\odot$ & 2.0-2.5 & 1.3-1.6 & 2.0-2.5  & 2.0-2.2 & 2.0 & 2.1-2.8 & 2.0-2.3 &  2.0-2.2 &  2.0-2.5 \\
\botrule
\end{tabular}}
\end{table}

We divide modern theoretical calculations of EoS of baryonic
matter into two groups.

{\it Microscopic calculations - baryon matter}. They are based on the
quantum many-body theories starting from realistic nuclear interactions, 
composed of two-body and three-body forces. 
Note that for Brueckner-Hartree-Fock calculations including nucleons and hyperons, 
$M_{\rm max}$ lies below $1.6~\msun$ while more massive NSs have been observed 
(see Sec.~\ref{sect:observations}): this is the ``hyperon puzzle''.

{\it Relativistic Mean Field calculations - baryon matter}. They are based on an 
effective relativistic lagrangian involving baryon and meson fields. The equations of motion
are solved in the mean-field approximation (RMF). Getting $M_{\rm max}>2~\msun$
for NSs with hyperonic cores is possible after a suitable adjustment of the parameters.

We considered different types of theories of quark matter cores in hybrid
stars. 

{\it Perturbative QCD}. The equations of the fundamental theory of quarks are solved 
perturbatively up to the second order in the strong coupling constant. However, it should 
be kept in mind that the convergence of this perturbative treatment is questionable for 
the conditions prevailing in the interior of a NS. 

{\it Effective theory of quark matter}. It relies on the effective
 Nambu-Jona-Lasinio (NJL) Lagrangian, whose solutions are obtained in 
 the mean-field approximation.

{\it Modified bag model of quark matter}. This model is based on the picture of 
 quarks confined inside a ``bag'', with significant corrections due to the 
 effective quark repulsion.
 
While microscopic calculations based on purely nucleonic matter can predict the existence of 
very massive NSs, getting $M_{\rm max}>2~\msun$ for hybrid stars requires a fine tuning of the 
model parameters: a phase transition at densities $\lesssim 2\rho_0$, a sufficiently strong vector 
repulsion between quarks and a small density jump at the baryon-matter - quark-matter 
interface. 

In the case of hypothetical family of ``twin compact stars'' -  a third family of compact stars, 
distinct from  WDs and NSs, and denser and more compact than NSs - the maximum mass 
$M_{\rm max}^{\rm twin}$ is usually lower than $2.0M_\odot$. However, 
with a fine tuning of the phase transition to quark matter and for a sufficiently stiff  EoS of quark 
matter, one can get ``in extremis'' a twin branch of hybrid stars  with a maximum mass 
$M_{\rm max}^{\rm twin}=2.0M_\odot$  (see, e.g., Ref.~\refcite{bla13}). On the other hand, results 
obtained using perturbative QCD for strange stars built of self-bound strange quark matter yield 
maximum masses in the range $2.0-2.7M_\odot$.
 
\begin{figure}[th]
\centerline{\psfig{file=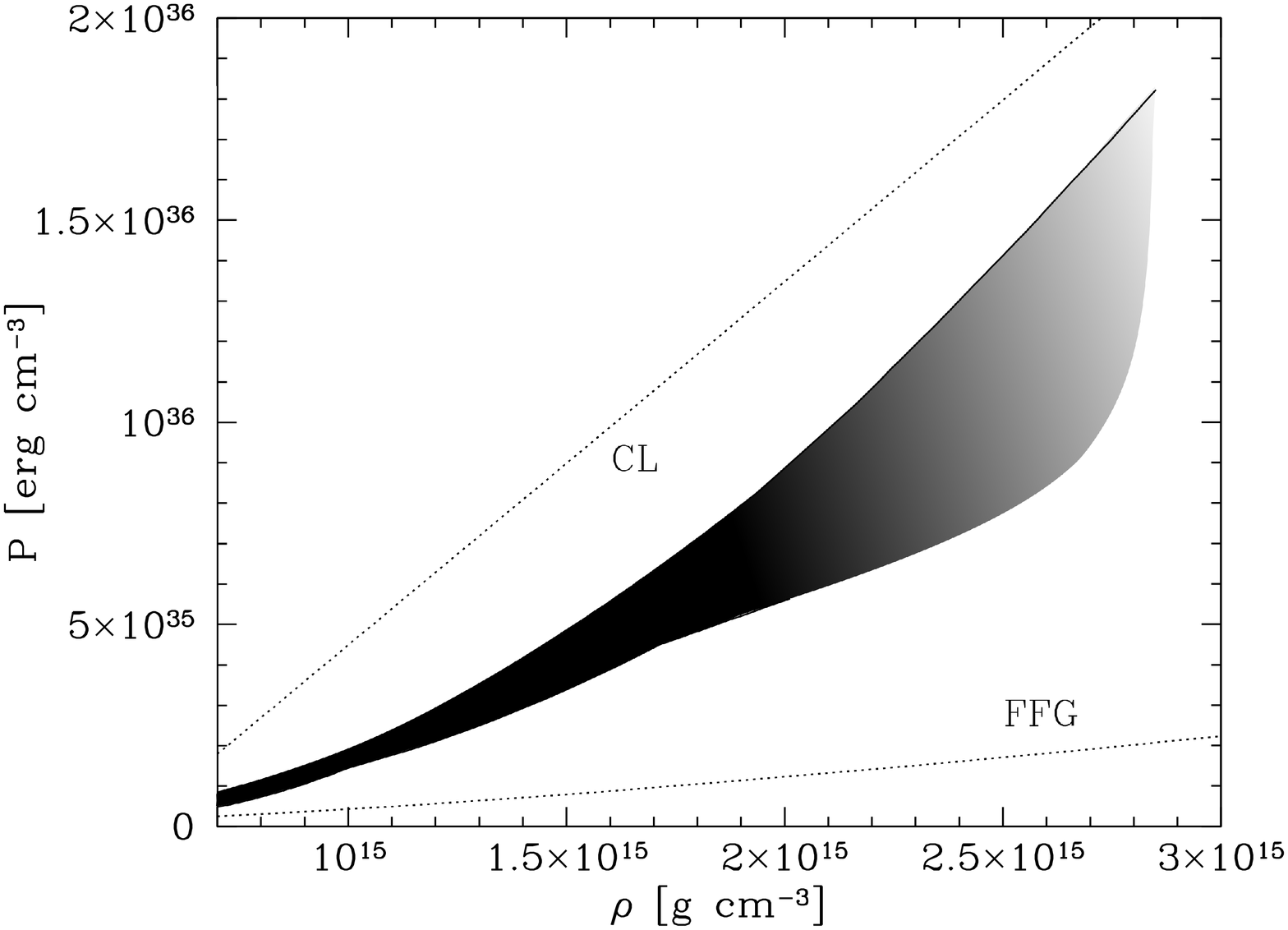,width=0.5\textwidth}\hskip 0.25cm\psfig{file=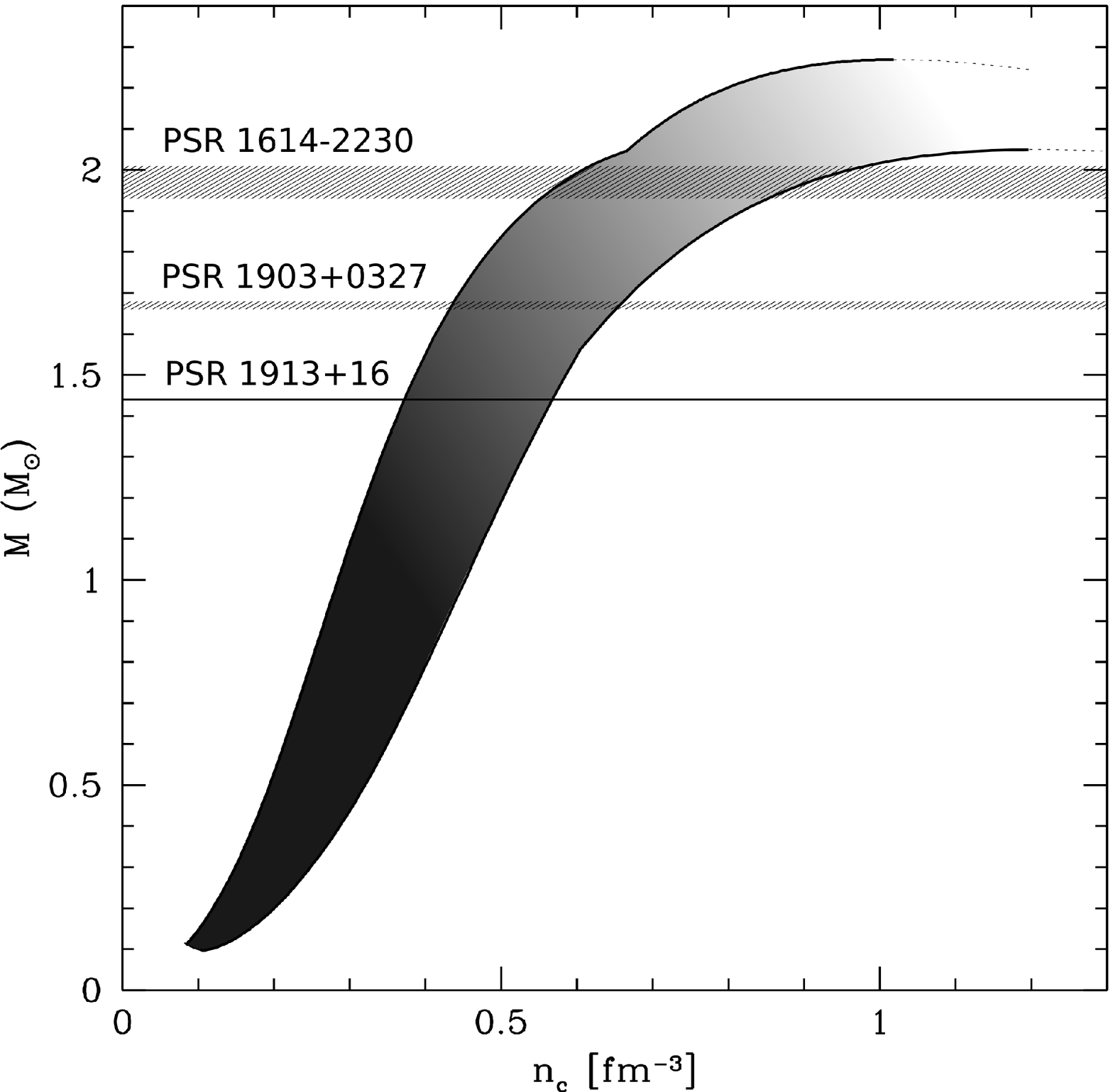,width=0.45\textwidth}}
\caption{Left panel: range of equations of state of dense matter (pressure $P$ versus mass density $\rho$), 
as predicted by various models and consistent with the existence of massive neutron stars.  The dotted lines 
labeled CL and FFG correspond to the causal limit and the free Fermi gas equations of state, respectively 
(see Sec.~\ref{sect:Mmax}).
Right panel: corresponding range of allowed masses $M$ for nonrotating neutron stars as a function of the 
central baryon number density $n_c$. The horizontal lines correspond to the precisely measured masses of 
three pulsars (see Sec.~\ref{sect:observations}).}
\label{fig:mass-radius}
\end{figure}

The current lack of knowledge of the EoS and the corresponding uncertainties 
in the predicted NS masses are illustrated in Fig.~\ref{fig:mass-radius}.

\section{Upper limits on the neutron-star maximum mass}
\label{sect:Mmax}

In view of all the uncertainties pertaining to the high-density part of the
EoS, the question arises as to whether meaningful constraints can be set
on the NS structure.

\begin{figure}[th]
\label{fig.Min}
\centerline{\psfig{file=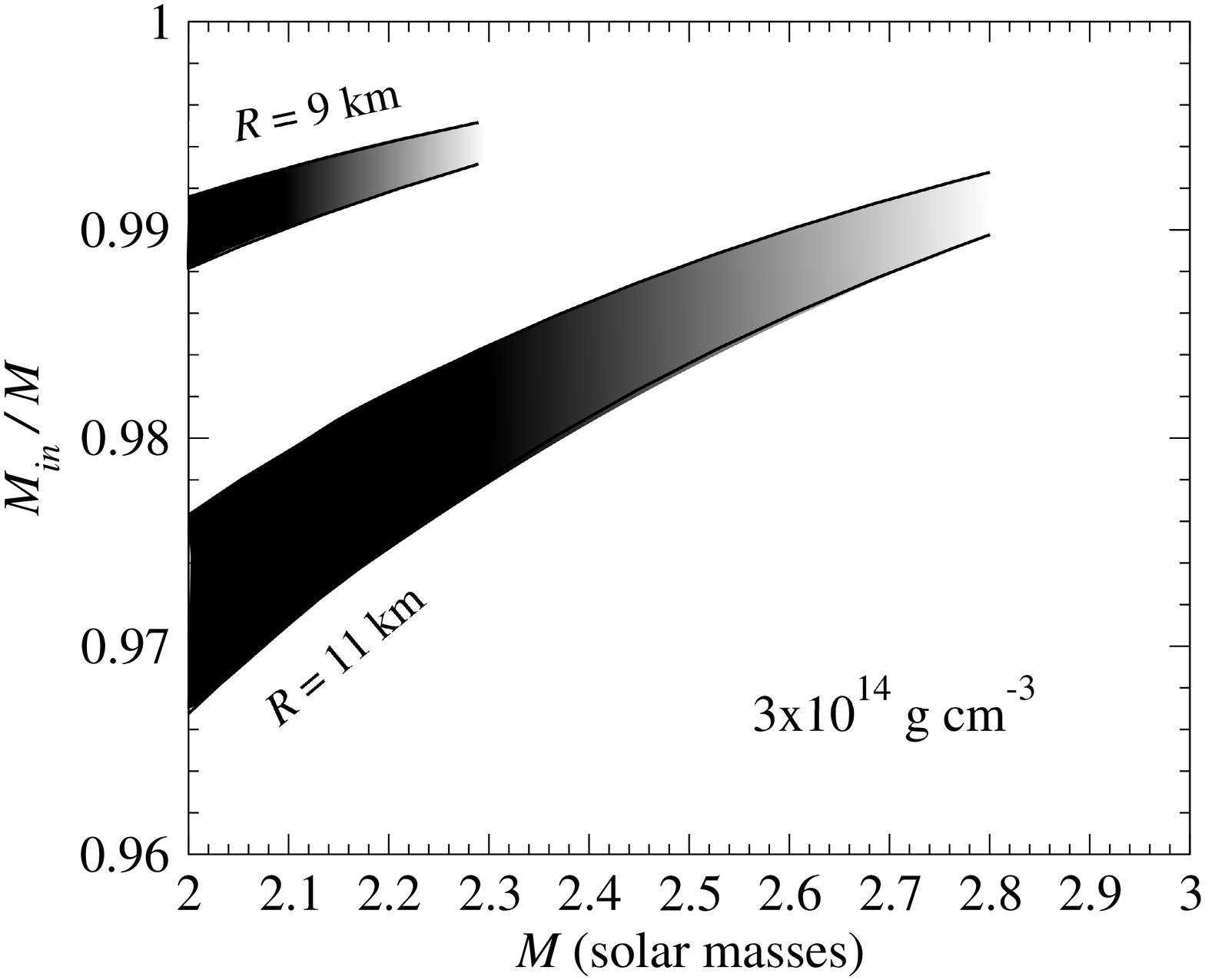,width=0.5\textwidth}\psfig{file=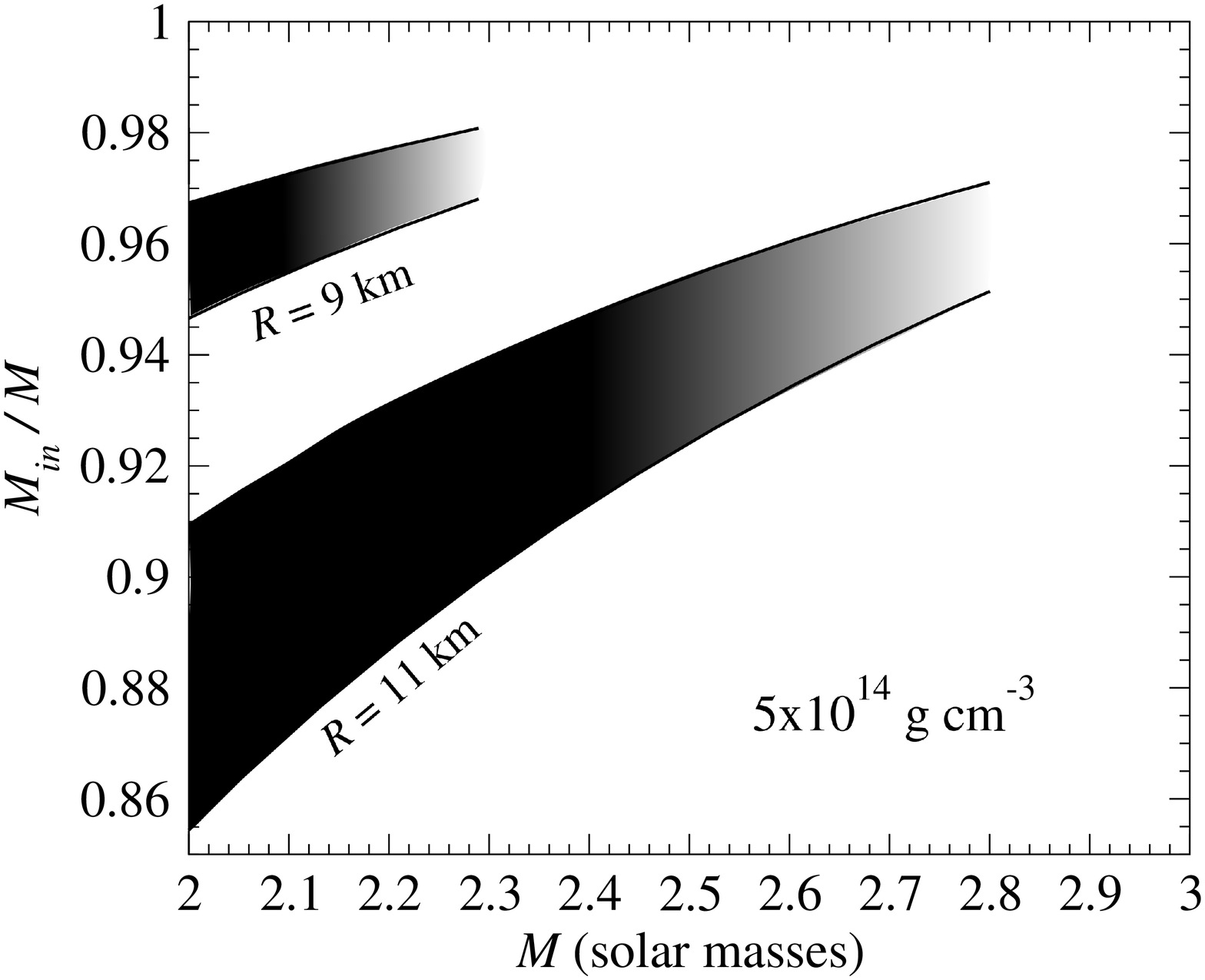,width=0.5\textwidth}}
\caption{Fractional mass $M_{\rm in}/M$ contained in the inner region of a static spherical NS of mass $M$ and radius $R$,
at density $\rho>\rho_\star$, for two different cases: $\rho_\star=3\times 10^{14}$ g~cm$^{-3}$ (left) and 
$\rho_\star =5\times 10^{14}$ g~cm$^{-3}$ (right).
The shaded areas reflect the uncertainties in the EoS~\cite{pgc11,pcgd12} at $\rho<\rho_\star$. Only the ranges of 
$M$ and $R$ allowed by the compactness constraint~\cite{bondi64,barr02} $r_{\rm g}/R \leq 6/8$ are shown. See the text for 
details.}
\end{figure}

Let us assume that the EoS of dense matter is reliably known up to some density
$\rho_\star$ for which $P=P_\star$. The mass of a static spherically symmetric NS
can thus be decomposed as $M=M_{\rm in}+M_{\rm out}$, where $M_{\rm in}$ ($M_{\rm
out}$) is the mass contained in the inner (outer) region of the NS at densities
above (below) $\rho_\star$. Typically, the density $\rho_\star$ lies in the range
between $\sim \rho_0$ and $\sim 2\rho_0$. We have estimated the contribution of 
the inner region to the mass of a NS by integrating inwards Eqs.~(\ref{tov1}) and
(\ref{tov2}) from the stellar surface to the radial coordinate $r=r_\star$ where
$\rho(r_\star)=\rho_\star$, for a given mass $M$ and radius $R$. For this purpose, 
we have use a set of unified EoSs that treat consistently both homogeneous and 
inhomogeneous phases~\cite{pgc11,pcgd12}. Note that the mass $M$ and radius $R$ are not 
completely arbitrary. In particular, the compactness $r_g/R$ of a NS is limited. 
Let us first recall that general relativity alone requires $r_g/R<1$: for a given 
radius $R$, the mass $M$ must thus be lower than $R c^2/(2G)$. The condition 
that the pressure at the center of the star should remain finite in order to prevent
the star from collapsing, translates to~\cite{buch59} $r_{\rm g}/R \leq 8/9 \simeq 0.889$.
The so called dominant energy condition~\cite{wald84} that the speed of energy flow cannot 
exceed the speed of light (i.e., $\rho(r) c^2\geq P(r)$) leads to the more stringent 
constraint~\cite{bondi64,barr02} $r_{\rm g}/R \leq 3/4 \simeq 0.75$. 
The mass $M_{\rm in}=\mathcal{M}(r_\star)$ is plotted in Fig.~\ref{fig.Min} for different 
radii and for the corresponding range of allowed NS masses, focusing on massive NSs. As 
shown on this figure, the more compact a NS is, the smaller is the contribution of the 
outer region to the stellar mass. 

Let us suppose for simplicity that $M\simeq M_{\rm in}$ and that the inner region is made of incompressible matter
at density $\rho_\star$. The following upper bound on the NS mass can thus be obtained~\cite{hlc75}
\begin{equation}
M\leq \frac{c^3}{G^{3/2}}
\left(\frac{3}{32\pi\rho_\star}\right)^{1/2}\left[1-\frac{1}{9}
\left(\frac{1+3\sigma_\star}{1+\sigma_\star}\right)^2\right]
\end{equation}
where $\sigma_\star=P_\star/(\rho_\star c^2)$. Since typically $\sigma_\star\ll 1$, we arrive at the following
estimate of the NS maximum mass
\begin{equation}
\label{struct-Mmax.inc}
M^{\rm inc}_{\rm max}\approx
5.09~\msun~\left(5\times 10^{14}~
\mathrm{g~cm^{-3}}\over \rho_\star\right)^{1\over 2}.
\end{equation}

Note that Eq.~(\ref{struct-Mmax.inc}) is a maximum maximorum since the assumption
of incompressible matter violates causality and special relativity.
A more stringent upper limit on the NS mass can be obtained by assuming that at
densities above $\rho_\star$, the EoS is the stiffest possible~\cite{zel62,nau73,rhru74,mal75,bre76,hlc75,har78,lat90,kal96,kor97,sag12}, 
with $dP/d{\rho}=c^2$:
\begin{equation}\label{eos.stiff}
P(\rho\geq \rho_\star) = c^2 (\rho-\rho_\star) + P_\star\, .
\end{equation}
Such an EoS is generally referred to as a causal limit (CL) EoS. For $\rho_\star \le 2\rho_0$, we get within a percent~\cite{hlc75,har78,wit84}
\begin{equation}
M^{\rm CL}_{\rm max} \approx  3.0~\msun~\left(5\times 10^{14}~
\mathrm{g~cm^{-3}}\over \rho_\star\right)^{1\over 2}.
\end{equation}
Setting $\rho_\star = 5\times 10^{14~}\;\mdens$, we obtain a rather
conservative upper bound on $M_{\rm max}<3\;\msun$ for nonrotating NSs, because we
are convinced that at $2\rho_0$ the speed of sound is lower than $c$.

\section{Effect of rotation on the maximum mass}
\label{sect:Mmax.rot}
Rotation increases the maximum mass of NSs because the centrifugal force acts
against gravity. We will consider two different cases: (i) rigidly rotating NSs,
and (ii) differentially rotating NSs.
\subsection{Rigid rotation}
\label{sect:Mmax.rigid}
In general relativity, a rigidly rotating star corresponds to stellar matter elements
moving around the rotation axis with a constant angular frequency $\Omega$, as measured by a distant observer
(see, e.g., Refs.~\refcite{haen07,Sterg2003} for a review of rotating NSs in general relativity).
In the perfect fluid approximation, rigidly rotating stationary configurations of NSs are axially
symmetric~\cite{Sterg2003,haen07}. Each configuration $\cal C$ can be characterized by two parameters, for example the
central density $\rho_{\rm c}$, and $\Omega$. The maximum mass of rigidly rotating NSs will be denoted by
$M_{\rm max}^{\rm rot(R)}$, where the superscript "R" is to remind that rotation is rigid. 
This configuration is not necessarily stable. Requiring the stability against axially symmetric perturbations leads
to an upper limit $\Omega_{\rm max}$ on the rotation frequency. It turns out that the mass of a NS rotating
at $\Omega_{\rm max}$ is very close to $M_{\rm max}^{\rm rot(R)}$. For realistic EoSs of dense matter, we have the
approximate relation $M_{\rm max}^{\rm rot(R)}\simeq 1.2\; M_{\rm max}^{\rm stat}$, where $M_{\rm max}^{\rm stat}$ is
the maximum mass of nonrotating (static and spherical) NSs.
For $\Omega<0.5\;\Omega_{\rm max}$, the maximum mass approximately increases as~\cite{haen07}
\begin{equation}\label{eq:Mmax.Omega}
M_{\rm max}^{\rm rot(R)}(\Omega)\simeq M_{\rm max}^{\rm stat}\left[ 1+0.2\;
\left(\Omega/\Omega_0\right)^2\right]\, ,
\end{equation}
where $\Omega_0=\sqrt{G M_{\rm max}^{\rm stat}/\left( R^{\rm stat}_{\max}\right)^3}$, $M_{\rm max}^{\rm stat}$ and
$R_{\rm max}^{\rm stat}$ being the maximum mass and corresponding radius of static NSs.
Setting $M=2M_\odot$ and $R=10~$km and using Eq.~(\ref{eq:Mmax.Omega}) we find that rotation increases the maximum mass
by $\sim 3\%$ only for PSR J1748$-$2446, whose frequency $f=\Omega/(2\pi)=716$~Hz is the highest measured.
For hypothetical bare quark stars built exclusively of
self-bound quark matter the effect of rigid rotation is significantly stronger, and
$M_{\rm max}^{\rm rot(R)}[{\rm QS}]\simeq 1.4\; M_{\rm max}^{\rm stat}[{\rm QS}]$~\cite{haen07}. The reason is
that the outer layers of bare quark stars are much more massive than those of baryonic stars.

\subsection{Differential  rotation}
\label{sect:Mmax.diff}
Let us consider the more general case of stationary and axially symmetric differentially rotating
NSs. Differential rotation  means that the angular frequency $\Omega$ depends on the distance $\varpi$
from the rotation axis (this is the only dependence allowed for stationary configurations). Hot
newly born NSs and the compact objects formed from the coalescence of two NSs in a binary system,
are expected to be differentially rotating because the associated dynamical time scales are too short
to allow for the transport of the angular momentum within the stellar interior.

The maximum equilibrium value of $\Omega$ at the equator coincides with the mass shedding limit, also
called the Keplerian frequency $\Omega_{\rm K}$ (i.e. the orbital frequency on a circular orbit in the equatorial plane
 just above the equator). Let us consider differentially rotating configurations with equatorial frequency
$\Omega_{\rm   eq}<\Omega_{\rm K}$, and $\Omega(\varpi)$ increasing inward such that $\Omega(0)>\Omega_{\rm K}$.
The centrifugal force acting on any matter element is larger for such differentially rotating NSs than for NSs rigidly
rotating at the same equatorial frequency $\Omega_{\rm  eq}$. Therefore, the maximum allowed mass $M_{\rm max}^{\rm rot(D)}$
of differentially rotating NSs (with $\Omega(\varpi)$ monotonously increasing with decreasing $\varpi$) will be larger
than the maximum mass $M_{\rm max}^{\rm rot(R)}(\Omega_{\rm eq})$ of NSs rigidly rotating at $\Omega_{\rm eq}$. The actual
value of $M_{\rm max}^{\rm rot(D)}$ not only depends on the EoS, but also on the functional form of $\Omega(\varpi)$
(see Ref.~\refcite{Galeazzi2012} and references therein). For a given function $\Omega(\varpi)$, the ratio
$M_{\rm max}^{\rm rot(D)}/M_{\rm max}^{\rm stat}$ is higher for bare quark stars built of self-bound quark matter
than for ordinary NSs~\cite{Szkudlarek2012}. While differential rotation can lead to masses as high as $3\;\msun -4~\msun$,
such configurations are secularly unstable. Differentially rotating stars will relax into a stationary state of rigid
rotation on a timescale  determined by the dominating angular momentum transport mechanism in the stellar interior. If
the transport is due to shear viscosity, differential rotation of a NS with internal temperature $10^9~$K will be damped
in $\sim 100~$years~\cite{Shapiro2000}. On the other hand, internal magnetic fields, however small, can convey
angular momentum much more effectively due to magneto-rotational instabilities~\cite{BalbusHawley1991}. As a result,
differential rotation is dissipated in seconds~\cite{Duez2006}.

\subsection{Supermassive and hypermassive rotating neutron stars}
\label{sect:super.hyper}
It stems from the preceding sections that we can distinguish three different maximum masses of NSs: $M_{\rm max}^{\rm stat}$
for nonrotating configurations, $M_{\rm max}^{\rm rot(R)}$ for rigidly rotating configurations, and $M_{\rm max}^{\rm rot(D)}$
for differentially rotating configurations. These limiting (gravitational) masses correspond to baryon masses $M_{\rm
b,max}^{\rm stat}$, $M_{\rm b,max}^{\rm rot(R)}$, and $M_{\rm b,max}^{\rm rot(D)}$ respectively. For baryon masses
$M_{\rm b}>M_{\rm b,max}^{\rm rot(D)}$ dense matter collapses into a rotating black hole. Differentially rotating NS
with baryon masses in the range $M_{\rm b,max}^{\rm rot(R)}<M_{\rm b}<M_{\rm b,max}^{\rm rot(D)}$ can only exist for
seconds: such stars are called {\it hypermassive}. Hypermassive NSs formed in binary NS mergers are doomed to collapse
into rotating black holes. Rigidly rotating NSs with baryon masses $M_{\rm b,max}^{\rm stat}<M_{\rm b}<M_{\rm
b,max}^{\rm rot(R)}$ can exist in a (quasi)stationary state provided their rotation is sufficiently rapid. However,
due to the loss of angular momentum (via electromagnetic radiation for instance), such stars eventually collapse into black holes
below some finite critical value of $\Omega$. These stars are called {\it supermassive}.

\section{Observations}
\label{sect:observations}
The discovery of the first binary pulsar PSR 1913+16 by Hulse and Taylor~\cite{Hulse1975}
enabled the first precise determination of NS masses. The masses of some two dozen binary pulsars
and their NS-companions were measured during the next decades, using pulsar timing analysis. But
until 2008, PSR 1913+16 remained the most massive NSs with a measured mass $M^{\rm (obs)}_{\rm max}=1.42\pm0.06~{\rm M}_\odot $
in 1982~\cite{Taylor1982}, $M^{\rm (obs)}_{\rm max}=1.442\pm0.003~{\rm M}_\odot $ in 1984~\cite{WeisbergTaylor1984}
and $M^{\rm (obs)}_{\rm max}= 1.4408\pm 0.0003~{\rm M}_\odot $ in 2003~\cite{WeisbergTaylor2003}.
In 2008, the millisecond pulsar PSR 1903+0327 replaced the Hulse-Taylor pulsar as the most massive NS~\cite{Champion2008}.
According to the most recent analysis, the mass of this pulsar is $M^{\rm (obs)}_{\rm max}= 1.67\pm 0.02~{\rm M}_\odot $
(the error bars correspond to an astonishing 99.7\% confidence level, see Ref.~\refcite{Freire2011}).
The discovery of the binary millisecond pulsar PSR J1614$-$2230 was very fortunate: (a) this system exhibits
a nearly edge-on orbital orientation with respect to the observer (within better than one arc degree), (b) the companion star is
a WD with a relatively high mass $0.5~\msun$. Both of these features enabled a precise determination of
the pulsar mass, whose measured value is $M^{\rm (obs)}_{\rm max}=1.97\pm 0.04~{\rm M}_\odot$ \cite{Demorest2010}, the
most massive NS known so far.
\footnote{After submission of the manuscript of this review  a  measurement of $2.01\pm 0.04\msun$ of PSR J0348+0432 was 
officially presented by Antoniadis et al.\cite{Antoniadis2013}. It does not change the conclusions of the present paper.}
The properties of these three binary pulsars are summarized in Table~2.
The general method used to measure NS masses and its application to the three binary pulsars are discussed in the
following sections.

\begin{table}[pt]
\tbl{Properties of the most massive binary pulsars. See the main text for explanations.}
{\begin{tabular}{@{}cccccccc@{}}\toprule
System &$m_{\rm P}(\msol)$&$m_{\rm C}(\msol)$  &$\pb$(d) & $P_{\rm s}$(ms) & $e$&  type &  discovery  \\
\hline
 B1913+16 & 1.44 & 1.39 &0.323 & 59.0 & 0.617 & NS-NS & 1974\cite{Hulse1975} \\
J1903+0327 & 1.67 & 1.05 & 95.17 & 2.15 & 0.437 & NS-MS & 2008\cite{Champion2008} \\
J1614-2230 & 1.97 & 0.5 & 8.7 & 3.15 & $1.3\times 10^{-6}$ & NS-WD & 2010\cite{Demorest2010} \\\botrule
\end{tabular}}
\end{table}

 \subsection{Precise measurements of neutron-star masses}
 \label{sect:M.observ}

\begin{figure}[th]
\centerline{\psfig{file=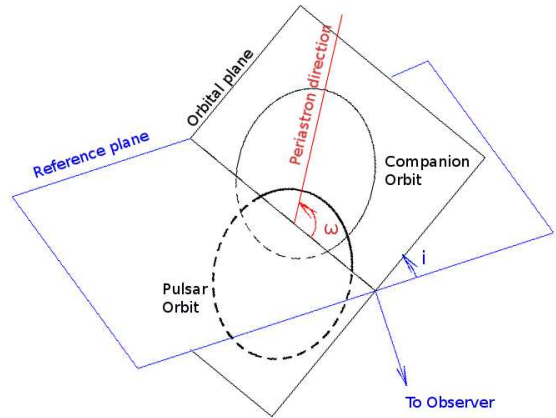,width=0.7\textwidth}}
\caption{Orbital parameters of a binary pulsar.}
\label{fig:orbit}
\end{figure}

The most accurate measurements of NS masses are based on observations of pulsars in binary systems.
The shift in the times of arrival (TOAs) of the pulses allows the determination of the pulsar's 
radial velocity (i.e., the velocity component along the direction to the observer as shown in Fig.~\ref{fig:orbit}),
as well as the parameters describing the pulsar's orbit. The orbital parameters are obtained assuming
Newtonian gravity and using Kepler's law to describe the orbital motion. Five Keplerian parameters
can be measured in binary systems where one star is observed as a pulsar. Three of them are connected with
the size and shape of the orbit: the binary orbital period $\pb$, the eccentricity of the orbit $e$ and the quantity
$x=(a\sin i)/c$ where $a$ is the semi-major axis and $i$ is the inclination angle of the orbit to
the line of sight, see Fig.~\ref{fig:orbit}. The two remaining parameters describe the orientation of the orbit
with respect to the observer: the longitude $\omega$ of the periastron and the reference time $T_0$ for the
orbiting pulsar defined by the time of periastron passage (see Fig.~\ref{fig:orbit}). In the Newtonian theory of binary motion for point-like masses,
these Keplerian parameters do not change in time in a local reference frame. However, due to the proper motion of the binary
system with respect to the Earth, the values of $\dot x$, $\od$, $\dot\pb$ may not be negligible.
The measurement of the five Keplerian parameters does not allow the determination of all the properties of the binary system.
Instead, we obtain two independent equations for four unknowns: the mass $m_{\rm P}$ of the pulsar, the mass $m_{\rm C}$ of the companion,
$a$ and $\sin i$.
One of these equations is simply $x=(a\sin i)/c$. The second equation arises from Kepler's laws and can be expressed
in terms of the mass function $f=(m_{\rm C}^3\sin^3 i)/M^2$ with $M=m_{\rm P}+m_{\rm C}$, as
\begin{equation}
f=\frac{4\pi^2}{\pb^2}x^3\, .
\end{equation}
The mass function is very useful in the analysis of binary systems since it provides a lower limit for the
companion mass obtained for $\sin i =1$.

The accurate description of tight binary systems containing compact objects requires the application of general
relativity. The deviations from Newtonian theory can be characterized by a few so called post-Keplerian parameters.
Some of these parameters directly describe the differences between the Keplerian orbit, which does not change in time,
and the general relativistic orbit, which does.

Due to the emission of gravitational waves, the binary system loses energy and orbital angular momentum. As a result, the period $\pb$
decreases and the orbit shrinks. This effect, usually characterized by a shift in the times of periastron passage, can be
accurately measured by a long-term monitoring of the binary system~\cite{Weisberg2010}. The orbital shrinking is most easily
observed in compact binary systems (low $\pb$) with a large eccentricity, although $\dot\pb\neq0$ also for circular orbits.
In general relativity, the orbit of a binary system is not closed: the major axis slowly rotates in the orbital plane. This
precession of the orbit leads to a secular variation $\od$ of the periastron longitude, which is best observed for tight
and highly eccentric binaries. The Doppler effect quadratic in the pulsar's velocity and the gravitational redshift in the field of the
companion can be characterized by a parameter $\gamma$, which depends on the masses $m_{\rm P}$ and $m_{\rm C}$. Since
$\gamma \propto e$, this parameter is most easily measured in highly-eccentric binaries.
As the pulsar's signals propagate through the curved space-time near the massive companion, they experience a gravitational delay.
The closer to the companion along the line of sight the pulsar is, the longer is the Shapiro delay. This effect is most pronounced
when the orbit is oriented edge-on (perpendicularly to the plane of the sky so that $\sin i = 1$) and when $m_{\rm C}$ is large.
Under these favorable circumstances, one can determine two parameters characterizing the Shapiro delay: its ``range'' $r\equiv G m_{\rm C}/c^3$ and
its ``shape'' $s\equiv \sin i$.

The post-Keplerian parameters can be expressed in terms of the Keplerian ones and the masses of the pulsar and its
companion~\cite{blandford76}:
\begin{eqnarray}
  \od &=& 3 \left( \frac{\pb}{2\pi} \right)^{-5/3}
  (T_\odot M)^{2/3} (1 - e^2)^{-1}\, ,\label{eq:dotom}\\
  \gamma &=& e \left(\frac{\pb}{2\pi}\right)^{1/3}
  T_\odot^{2/3}M^{-4/3} m_{\rm C} (m_\mathrm{p} + 2m_{\rm C})\, ,\\
 \dot \pb &=& -\frac{192\pi}{5}
  \left( \frac{\pb}{2\pi} \right)^{-5/3}
  \left( 1 + \frac{73}{24} e^2 + \frac{37}{96} e^4 \right)
  \left( 1 - e^2 \right)^{-7/2}
  T_\odot^{5/3} \frac{m_{\rm P} m_{\rm C}}{M^{1/3}}\, ,\label{eq:pdot}\\
  r &=& T_\odot m_{\rm C}\, ,
  \label{equ:r}\\
  s &=& x \left( \frac{\pb}{2\pi} \right)^{-2/3}T_\odot^{-1/3} M^{2/3} m_{\rm C}^{-1} \label{eq:s}\, ,
\end{eqnarray}
where $T_\odot\equiv G M_\odot/c^3 = 4.925490947\mathrm{\ \mu s}$, and $m_{\rm P}$, $m_{\rm C}$ are masses of pulsar and
companion in solar unit $\msun$.

\begin{figure}[th]
\centerline{\psfig{file=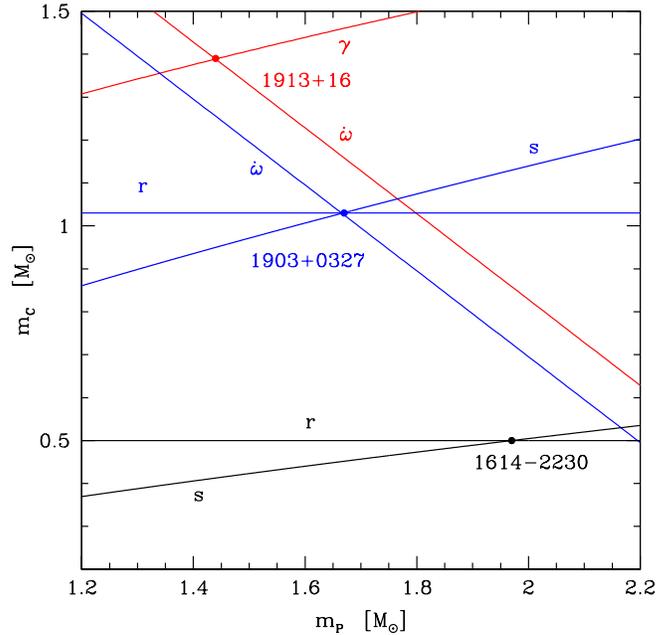,width=0.7\textwidth}}
\caption{Masses of the pulsar and its companion of three binary systems, as determined by the measurement
of the post-Keplerian parameters $\od$, $\gamma$. $s$ and $r$. In the case of PSR 1903+0327, 3 parameters
were used to determine $m_{\rm P}$ at 99.7\% confidence level.}
\label{fig:msum}
\end{figure}

Given the precisely measured Keplerian parameters, one can determine all the coefficients in Eqs.~(\ref{eq:dotom}-\ref{eq:s}) with
only two unknowns: the masses $m_{\rm P}$ and $m_{\rm C}$. From a measurement of just two  post-Keplerian parameters one can therefore
solve for the two masses and determine all the parameters of the binary system, including its orientation (the inclination angle $i$). If three
(or more) post-Keplerian parameters are measured, the system of Eqs.~(\ref{eq:dotom}-\ref{eq:s}) is overdetermined, thus offering the opportunity to
test the theory of gravitation (see, e.g., Ref.~\refcite{Stairs2003}). The relation between $m_P$ and $m_C$ resulting from the measurement of
post-Keplerian parameters for three discussed pulsars is presented in Fig.~\ref{fig:msum}.

\subsection{PSR 1913+16}

The Hulse-Taylor pulsar was the first radio pulsar in a double NS system for which the relativistic corrections to the Keplerian motion were observed.
The first measured parameter was the advance $\od$ of the longitude of the periastron~\cite{Taylor1976}. A recent analysis of observational
data yields $\od= 4.226598(5)$ with a relative error of $10^{-6}$. Using Eq.(\ref{eq:dotom}) leads to a determination of the total mass of
the system: $M=2.828378(7)\msol$~\cite{Weisberg2010}.
The second measured parameter was the parameter $\gamma$~\cite{Taylor1979,Taylor1982,Taylor1989}. As discussed in the previous section,
the measurements of both $\od$ and $\gamma$ allows the determination of the individual NS masses with an accuracy of the order
of $10^{-4}$. The third measured parameter is $\dot\pb$ - the rate of decrease of the orbital period, given by the Eq.~(\ref{eq:pdot}).
The value predicted by general relativity is $\dot\pb^{\rm GR}=-2.4025\times 10^{-12}$, while the observed value is $\dot\pb=-2.423(1)\times 10^{-12}$. Actually, the value of $\dot\pb$ that is measured includes systematic effects caused by the relative acceleration of the solar
system with respect to the binary system~\cite{Damour1991}. Recent estimate yields for this kinematic contribution the
value~\cite{Weisberg2010} $\Delta\dot\pb=-0.027\pm0.005\times 10^{-12}$. Correcting for this effect, the observed value of $\dot\pb$
is in excellent agreement with Einstein's theory of general relativity.

\subsection{PSR 1903+0327}

The system PSR 1903+0327 is unusual in many respects, including the companion type (a main sequence star of $1\msun$ mass) and
a highly eccentric orbit with $e=0.44$ (the only Galactic millisecond pulsar of this kind). These properties challenge the evolutionary
scenarios for the formation of millisecond pulsars~\cite{Champion2008,Liu2009,Portegies2011,Bejger2011,Freire2011}.
A precise determination of $\dot\omega$ is hampered by the change of the orbital orientation with respect to the observer due to
the proper motion of the system~\cite{Kopeikin1996}. This effect, which could be about seven times larger than the accuracy of the
$\dot\omega$ measurement, is the main source of uncertainty in estimation of the total mass of the system.
Assuming that the properties of the companion are similar to those of our Sun, the contribution to the periastron shift 
of the centrifugal flattening of the companion resulting from its rotation is estimated to be a few times smaller than the measured uncertainty of
$\od$~\cite{Freire2011}. However the rotation of the companion is not well (observationally) constrained and the spin-orbit contribution could be
significantly larger, even if we apply the recent limit $\le 66~\mathrm{km}/\mathrm{s}$ for the companion's rotational speed~\cite{Khargharia2012}.
The almost edge-on orientation of the orbit ($\sin i =0.97$) allows the determination of the two Shapiro parameters.
However, the analysis is complicated by the fact that the companion is a main-sequence star. For example, the change
of dispersion due to stellar wind could mimic a Shapiro delay. The analysis of dispersion measure as a function of the
orbital phase at different frequencies proved that this effect is negligible~\cite{Freire2011}.
The measurement of three post-Keplerian parameters has led to the value $m_{\rm P}=1.667\pm0.021\msun$ at  $99.7\%$ confidence
level~\cite{Freire2011}.

\subsection{PSR 1614-2230}

The detailed analysis of the binary orbit of the pulsar PSR 1614-2230 was published in 2010~\cite{Demorest2010}.
This system consist of a $2\msun$ pulsar and a helium-carbon-oxygen WD (with a mass of $0.5\msun$) on a
nearly circular edge-on orbit ($i=89.17^\circ$). The determination of the masses is based on the measurement of the
two Shapiro delay parameters $r$ and $s$. The NS mass is $1.97\msun \pm0.04\msun$  at $1\sigma$-accuracy
and $\pm0.11\msun$ at $3\sigma$~\cite{Demorest2010}. This accuracy of this measurement is unlikely to improve in the
near future since the Shapiro delay does not accumulate over time (contrary to $\dot\omega$ and $\dot\pb$).

\subsection{Other neutron-star mass measurements}

Historically, the mass measurements of the three pulsars PSR 1913+16, PSR 1903+0327 and PSR 1614$-$2230 reviewed in 
the previous sections, set the highest limit on the NS mass. The basic theory behind those measurements 
is general relativity, which seems to be the correct theory for describing gravitational interactions. Moreover, the extremely 
accurate analysis of radio pulses leads to a very precise determination of the relativistic parameters of the binary 
motion.

It should be mentioned, however, that there exists quite a large number of less precise measurement of NS masses with values 
of the order and even above $2\msun$~\cite{haen07}. These measurements mainly refer to NSs in X-ray binaries, where accretion, stellar
wind, possible filling of Roche lobe by the companion could all play an important role. For this reason, the error 
of these NS mass measurements is quite large, typically a few tenths of $\msun$ (see, e.g., Ref.~\refcite{haen07} for 
a discussion).

Recent observations suggest that the so called black-widow pulsar PSR B1957+20 might be a very massive NS~\cite{kbk11}.
The analysis of this system is based on the observations of a companion with a very low mass $\simeq 0.03\msun$.
The estimated mass of the pulsar is $2.4\msun$. But taking into account the possible systematic uncertainties leads to 
a lower limit of $1.7\msun$ for this pulsar's mass.

\section{Discussion and conclusion}
\label{sect:discussion.conclusion}

The maximum mass of a NS is a direct consequence of general relativity and depends on the EoS at
densities ranging from that of ordinary matter up to about $10\rho_0$. The EoS is well established for $\rho\lesssim \rho_0$,
reasonably well for $\rho_0<\rho\lesssim 2\rho_0$, but it is very uncertain in the range
$2\rho_0\lesssim \rho< 10\;\rho_0$. Alas, the value of the maximum mass is to a large extent
determined by the high-density part of the EoS. The ``theoretical uncertainty'' reflects
our lack of a precise knowledge of the strong interactions of the dense matter constituents, and
stems also from deficiencies and uncontrollable approximations of the many-body theory of the
strongly interacting system under consideration. According to different calculations, the
maximum mass of spherical nonrotating NSs is predicted to lie in the range
$1.5~\msun \lesssim M_{\rm max}\lesssim 2.5~\msun$. Rotation increases the maximum mass. However, this increase
amounts to $\sim 3\%$ only for PSR J1748$-$2446, the most rapidly spinning pulsar known. Higher masses
could be reached in differentially rotating NSs, but such configurations are secularly unstable
on timescales of seconds.

NS masses have been precisely measured for some binary pulsars. Until very recently, the largest precisely
measured NS mass was $M^{\rm (obs)}_{\rm max}=1.97\pm 0.04~\msun$ for PSR J1614$-$2230~\cite{Demorest2010}.
However, at the time of writing the precise measurement of the mass of pulsar PSR J0348+0432~\cite{Antoniadis2013} 
may set a new limit for NS masses: $M^{\rm (obs)}_{\rm max}=2.01\pm 0.04~\msun$.
This mass is sufficiently high to put quite strong constraints on the poorly known EoS of dense matter at densities
$\rho>4\rho_0$. However, it still remains compatible with a large class of models. On the other hand, this measured
mass, which is about three times larger than the maximum mass of a star made of an ideal neutron Fermi gas, is a
clear observational indication of the dominating role of strong interactions in NSs. In contrast, the 
maximum mass of a WD can be fairly accurately using the EoS of an almost ideal electron Fermi gas.

Future measurements of NS masses substantially higher than $2.5~\msun$ would  be a
real challenge for modern theory of dense matter. For the time being, we find it
reasonable to assume that cold matter at densities $\rho<5\times 10^{14}~\mdens$ is
nucleonic, and that for such  densities its EoS is reasonably well known. Then, as
we have seen, the condition that the sound speed for $\rho>5\times 10^{14}~\mdens$ not
exceed $c$ implies an absolute upper bound on the NS mass of $3\;\msun$. Therefore, we 
conclude that the true maximum mass of NSs is between $2\;\msun$ and $3\;\msun$.

\section*{Acknowledgements}

The work of N.C. and A.F.F. was financially supported by FNRS (Belgium).
This work was partially supported by the Polish NCN grant no 2011/01/B/ST9/04838.

\end{document}